\documentclass[aps,pre,preprint,nopreprintnumbers,amssymb,amsmath,showpacs,showkeys,fleqn,eqsecnum,superscriptaddress]{revtex4-1}

\usepackage[final]{graphicx}
\usepackage{amssymb}
\usepackage{amsmath}
\usepackage{dcolumn}

\begin{document}

\title{Fast Converging Path Integrals for Time-Dependent Potentials I: Recursive Calculation of Short-Time Expansion of the Propagator}
\author{Antun Bala\v{z}}\email[E-mail: ]{antun@ipb.ac.rs}
\affiliation{Scientific Computing Laboratory, Institute of Physics Belgrade, University of Belgrade, Pregrevica 118, 11080 Belgrade, Serbia}
\homepage[Home page: ]{http://www.scl.rs/}
\affiliation{Department of Physics, Faculty of Sciences, University of Novi Sad, Trg Dositeja Obradovi\' ca 4, 21000 Novi Sad, Serbia}
\author{Ivana Vidanovi\'c}
\author{Aleksandar Bogojevi\'c}
\author{Aleksandar Beli\'c}
\affiliation{Scientific Computing Laboratory, Institute of Physics Belgrade, University of Belgrade, Pregrevica 118, 11080 Belgrade, Serbia}
\homepage[Home page: ]{http://www.scl.rs/}
\author{Axel Pelster}
\affiliation{Fachbereich Physik, Universit\" at Duisburg-Essen, Lotharstra\ss e 1, 47048 Duisburg, Germany}

\begin{abstract}
In this and subsequent paper \cite{fcpitdp2} we develop a recursive approach for calculating the short-time expansion of the propagator for a general quantum system in a time-dependent potential to orders that have not yet been accessible before. To this end the propagator is expressed in terms of a discretized effective potential, for which we derive and analytically solve a set of efficient recursion relations. Such a discretized effective potential can be used to substantially speed up numerical Monte Carlo simulations for path integrals, or to set up various analytic approximation techniques to study properties of quantum systems in time-dependent potentials. The analytically derived results are numerically verified by treating several simple models.
\end{abstract}

\maketitle

\section{Introduction}
\label{sec:intro}
Studying quantum systems in time-dependent potentials represents a fundamental 
problem which emer\-ges in many areas of physics. Even if the Hamiltonian of the system itself does not explicitly depend on time, such situations 
naturally occur in the presence of time-dependent external fields. Another important example are fast-rotating systems, where 
the rotation introduces an explicit time dependence of the respective variables in the co-rotating 
frame. This includes modern experiments with rotating ultracold atoms and Bose-Einstein condensates \cite{dalibard2004, fetter2005, abdullaevpla, pelster1,theodorakis, alon, pelster2, mason2009}, studies of vortices
 \cite{aftalion1,aftalion2}, as well as non-stationary optical lattices \cite{inguscio, ketterle1,ketterle2}. In the latter case counter-propagating laser beams are not perfectly modulated, 
so they produce an effectively moving optical lattice. An explicit 
treatment of time-dependent problems is also often required in nuclear and molecular physics \cite{deumens}, and in studies of entanglement 
phenomena \cite{boness,eisler,bellomo}. 
Note that some systems can also be treated effectively with a time-dependent potential, which actually represents 
higher nonlinear terms in the wave function. A prominent example is the Gross-Pitaevskii equation, where the nonlinear term is treated within the split-step Crank-Nicolson method
as a time-dependent potential which is updated after each time step \cite{muru}.

A time-dependent formalism has been developed in the framework of various theoretical approaches, based on the time-dependent Schr\" odinger equation. The well-known time-dependent semiclassical approximation is formulated by the Van Vleck propagator, and provides wide-range of important results, including generalizations to the time-dependent variational perturbation theory. Time-dependent variational principle, formulated by Dirac, represents basis for the time-dependent self-consistent fields method \cite{heller}. Another general method is time-dependent perturbation theory,  which is also applied to derive a number of other approaches, including time-dependent scattering theory, linear response theory (Kubo formula) and Fermi's golden rule. Path integral formalism \cite{kleinertbook} also provides general framework to study dynamics of time-dependent quantum systems. In addition to these general methods, time-dependent variants of many highly specialized methods are also developed. As important examples, let us mention time-dependent Density Matrix Renormalization Group (DMRG) \cite{schollwoeck,hallberg}, the Density Functional Theory (DFT) \cite{runge,onida}, and the Density Matrix Functional Theory (DMFT) \cite{pernal}.

In numerical approaches for time-dependent systems, partial differential equation representation of the time-dependent Schr\" odinger equation can be solved using the discretization of the time domain through the finite difference method. In this case, the specific techniques include both explicit and implicit schemes, with the commonly used Crank-Nicolson scheme \cite{crank}. Other major numerical approaches include molecular dynamics, as well as the split-operator method, which is based on a space-grid calculation both without  \cite{chinkrotscheck,hernandez}
and with Monte Carlo \cite{ciftja,sakkos}. This method is mainly used in its second-order variant in the time of propagation for time-dependent potentials, where no change is required in comparison with the time-independent case. However, also higher order split-operator schemes have been derived, including fourth \cite{bandrauk, blanes, chinchen, bayepla} and higher-order expansions \cite{omelyanpre, omelyan, bayepre, zillich}. We also mention important counterexamples, like world-line Monte Carlo \cite{evertz, prokofev-jetp, sylj}, where discretization errors are completely removed \cite{beard}, or open systems where the main contribution to the systematic error is due to retardation effects which cannot be dressed into the form of simple potentials \cite{breuer}.

In this and subsequent paper \cite{fcpitdp2}, we extend the earlier established approach in Refs.~\cite{prl-speedup, prb-speedup, pla-manybody, balazpre} for obtaining a high-order short-time expansion of transition amplitudes of
time-independent potentials to the important time-dependent case. This development allows a high-precision calculation of transition amplitudes, which is necessary for 
extracting properties of various 
quantum systems, such as partition functions. Note that individual transition
amplitudes can be accurately calculated using the lower-order effective actions at the expense of  increasing the number of Monte Carlo time-steps, 
which would just increase the needed computation time. Although the presented approach can be used for improving the efficiency for calculating such 
transition amplitudes, it is mainly developed  for
applications which require a large number of accurate transition amplitudes for further numerical calculations. Such situations occur, for instance, for determining
partition functions \cite{pla-manybody, danicapla} or for obtaining energy spectra 
with the method of diagonalizing the space-discretized evolution operator matrix \cite{diag1, diag2}. 
In order  to avoid an accumulation of numerical errors, so that
such calculations can be performed with the required accuracy, the transition amplitudes have to 
be known more precisely.

Our approach has already been successively used to study global and local properties of rotating Bose-Einstein condensates \cite{becpla}, where 
the availability of accurately calculated thermal states has allowed to calculate precisely 
the condensation temperature, the ground-state occupancy, and time-of-flight absorption pictures even in the delicate regime of a critical and an overcritical rotation. 
If one needs a large numbers of accurate 
energy eigenvalues and eigenstates, as in this case, higher-order effective actions become useful. The precise knowledge of the short-time expansion of the propagator might also be used 
to improve numerical studies of time evolution, especially in systems exhibiting nonlinearities \cite{chinkrotscheck}, where a time-dependent formalism is required. In addition to this, the presented approach can offer a significant improvement in studying dynamics of quantum systems within the Path Integral Monte Carlo calculations. The multilevel blocking method \cite{mak} might be used to address the inherent dynamical sign problem, while the required number of Trotter time slices may be significantly decreased, due to high order of convergence of short-time propagators derived here.

Thus, in spirit of Symanzik`s improved action programme in quantum field theory \cite{symanzik,fliegner},
we will introduce and analytically derive effective actions for time-dependent potentials 
which substantially speed up the
numerical calculation of transition amplitudes and other quantities for such quantum systems. This approach does not suffer from technical problems related to backwards diffusion in time, 
as observed and discussed
by Sheng \cite{sheng} and Suzuki \cite{suzuki}. In the split-operator method one has to resort to multiproduct expansions \cite{k1} in order to resolve this problem. In the higher-order 
effective action approach for time-dependent potentials, which is presented here, the convergence of transition amplitudes is always guaranteed in a natural way, as was shown conclusively
for the time-independent case in Ref.~\cite{pre-ideal}. While this approach has a number of benefits (systematic higher-order expansion, simple implementation of symbolic calculation of effective actions using recursive relations, unconditional convergence of transition amplitudes in the imaginary-time formalism, simple numerical implementation of the algorithm), it also has some disadvantages when compared to other short-time approaches. Most notably, higher-order effective action in this method contain higher spatial and time derivatives of the potential, and thus require the potential to be smooth function. Another possible disadvantage is increasing computational complexity of higher-order effective actions, which significantly grows with the order, and has to be appropriately chosen so as to minimize the computing time for the desired precision of numerical calculations.

The outline of this paper is as follows. In Sec.~\ref{sec:piformalism} we briefly review the underlying path-integral formalism
of quantum mechanics \cite{feynman,feynmanhibbs,feynmanstat,kleinertbook} for systems in time-dependent potentials in order to fix our notation. Afterwards, Sec.~\ref{sec:pi} presents a lowest-order path-integral
calculation of the short-time transition amplitude for time-dependent potentials in the single-particle one-dimensional case. Based on these lowest-order results and using Schr\" odinger equations for transition amplitudes of systems in time-dependent potentials derived in Sec.~\ref{sec:sch}, we develop in Sec.~\ref{sec:recursion} the systematic efficient recursive approach for one dimensional quantum systems. The analytically derived results are verified numerically with the help of several illustrative one-dimensional models in Sec.~\ref{sec:numver}.

The approach developed here to obtain transition amplitudes for time-dependent potentials is generalized in the second part of this paper \cite{fcpitdp2} to quantum systems with many degrees of freedom. In the next paper we also show how the developed imaginary-time formalism is transformed into a real-time one, and demonstrate its applicability by treating several models.

\section{Path-Integral Formalism for Systems with Time-Dependent Potentials}
\label{sec:piformalism}
We will consider a non-relativistic quantum multi-component system in $d$ spatial dimensions with a Hamilton operator which consists of the usual kinetic term and a time-dependent 
potential:
\begin{equation}
\hat H(\hat{\mathbf p}, \hat{\mathbf q}, t)=\sum_{i=1}^P\frac{\hat{\mathbf p}_{(i)}^2}{2M_{(i)}}+V(\hat{\mathbf q}, t)\, .
\end{equation}
Here $P$ stands for 
the number of particles, the $P\times d$ dimensional vectors $\mathbf q$ and $\mathbf p$ describe positions and momenta of all particles, and the parenthetic subscript 
$(i)$ denotes the corresponding quantity for particle $i$. Although in this paper we only consider single-particle systems, here we introduce many-body notation and use it in introductory parts of Sections \ref{sec:pi} and \ref{sec:sch}, as a preparation for the second part of this paper \cite{fcpitdp2}, where extensions of the presented approach to many-body systems and to the real-time formalism are developed.

Note that the potential $V(\mathbf q, t)$ of the system is allowed to depend on the positions of all particles, and, therefore, contains implicitly all types of interactions. In practical applications the potential $V$ usually contains, apart from the external potential, also two- and three-body interactions, but further many-body interactions can, in principle, be included as well.

The central object for studying the dynamics of such a quantum system within the path-integral formulation is the transition amplitude
\begin{equation}
\label{eq:AU}
A(\mathbf a,t_a; \mathbf b, t_b)=\langle \mathbf b, t_b|\hat U(t_a\to t_b)|\mathbf a, t_a\rangle\, .
\end{equation}
Here the vectors $\mathbf a$ and $\mathbf b$ describe the positions of all particles at the initial and final time $t_a$ and $t_b$, $|\mathbf a, t_a\rangle$ and 
$|\mathbf b, t_b\rangle$ 
denote the corresponding Hilbert-space states of the system, and
\begin{equation}
\label{eq:UT}
\hat U(t_a\to t_b)=\hat{T}\exp\left\{-\frac{i}{\hbar}\int_{t_a}^{t_b}dt\, \hat H(\hat{\mathbf p}, \hat{\mathbf q}, t) \right\}
\end{equation}
represents the evolution operator of the system describing its propagation from $t_a$ to $t_b$.
Here we assumed $t_a<t_b$ and introduced the standard time-ordering operator
\begin{equation}
\hat{T} \{\hat O(t)\hat O(t')\}= \left\{ \begin{array}{ll}
\hat O(t)\hat O(t'), & \textrm{if $t> t'$,}\\
\hat O(t')\hat O(t), & \textrm{otherwise.}
\end{array} \right. \, .
\end{equation}
The starting point in setting up the path-integral formalism is the completeness relation
\begin{equation}
\hspace*{-8mm}
\label{eq:complete}
A(\mathbf a,t_a; \mathbf b, t_b)=\int d\mathbf q_1\cdots \int d\mathbf q_{N-1} \, A(\mathbf a,t_a; \mathbf q_1, t_1)
\,A(\mathbf q_1,t_1; \mathbf q_2, t_2)\cdots A(\mathbf q_{N-1},t_{N-1}; \mathbf b, t_b)\, ,
\end{equation}
where $\varepsilon=(t_b-t_a)/N$ denotes the time-slice width, $t_n=t_a+n \varepsilon$ are discrete time steps, and the 
$P\times d$ dimensional vectors $\mathbf q_1,\ldots, \mathbf q_{N-1}$ 
describe positions of all particles at a given discrete time step which is specified by the non-parenthetic index. To leading order in $\varepsilon$, the short-time
transition amplitude reads
\begin{equation}
\label{eq:lowest}
A(\mathbf q_n, t_n; \mathbf q_{n+1},t_{n+1})
\approx \frac{1}{(2\pi\hbar i \varepsilon)^{Pd/2}}\exp \left\{\frac{i}{\hbar} S^{(1)}(\mathbf q_n, t_n; \mathbf q_{n+1},t_{n+1}) \right\}\, ,
\end{equation}
where the naive discretized action $S^{(1)}$ is usually expressed as
\begin{equation}
S^{(1)}(\mathbf q_n, t_n; \mathbf q_{n+1},t_{n+1})=\varepsilon \left\{\frac{1}{2}\left(\frac{\boldsymbol{\delta}_n}{\varepsilon}\right)^2\hspace{-2mm}-V(\mathbf x_n, \tau_n)\right\} .
\label{eq:S1}
\end{equation}
Here we introduced the discretized velocity $\boldsymbol{\delta}_n=\mathbf q_{n+1}-\mathbf q_n$, and rescaled the coordinates so that the mass of all particles is equal to unity. 
The potential $V$ is evaluated at the mid-point coordinate $\mathbf x_n=(\mathbf q_n+\mathbf q_{n+1})/2$ and at the mid-point time $\tau_n=(t_n+t_{n+1})/2$. Eq.~(\ref{eq:S1}) is correct to 
order $\varepsilon\sim 1/N$ and, therefore, after substitution to Eq.~(\ref{eq:lowest}), leads to errors of the order $O(1/N^2)$ for discretized short-time 
transition amplitudes. Note that the normalization 
factor $\sim 1/\varepsilon^{Pd/2}$ in Eq.~(\ref{eq:lowest}) does not affect the $N$-scaling of errors, since short-time amplitudes will be inserted into the completeness relation (\ref{eq:complete}), 
where this normalization factor will be added to the normalization of the long-time transition
amplitude. However, due to the fact that Eq.~(\ref{eq:complete}) contains a product of $N$ short-time transition
amplitudes, the deviation of the obtained discrete transition amplitude from the corresponding continuum result will be of the order $N\cdot O(1/N^2)=O(1/N)$. For this reason the naive discretized 
action is designated by $S^{(1)}$. 

In the limit $N\to\infty$ we recover the continuous transition amplitude, which leads to the formal coordinate-space path-integral expression
\begin{equation}
\label{eq:pi}
A(\mathbf a,t_a; \mathbf b, t_b)=\int_{\mathbf q(t_a)=\mathbf a}^{\mathbf q(t_b)=\mathbf b}  {\cal D} \mathbf q(t)\, \exp \left\{\frac{i}{\hbar} S[\mathbf q]\right\}\, ,
\end{equation}
where the integration is defined over all possible trajectories $\mathbf q(t)$ through the discretization process described above. In this equation, the action $S$ for a given 
trajectory $\mathbf q(t)$ is defined as usual:
\begin{equation}
\label{eq:Sdef}
S[\mathbf q]=\int_{t_a}^{t_b} dt\, \left\{ \frac{1}{2}\dot{\mathbf q}^2(t)-V(\mathbf q(t), t)\right\}\, .
\end{equation}
The outlined derivation represents the basis for the path-integral 
formulation of quantum mechanics \cite{feynman,feynmanhibbs,feynmanstat,kleinertbook} as well as for the numerical calculation of path integrals. The described discretization 
procedure is most straightforwardly numerically implemented  by the Path-Integral Monte Carlo approach \cite{ceperley}.

Note that the $O(1/N)$ convergence can be also achie\-ved with other choices of space-time points at which we evaluate the potential in Eq.~(\ref{eq:S1}). For example, 
the left or the right prescription $\mathbf x_n=\mathbf q_n, \tau_n=t_n$ or $\mathbf x_n=\mathbf q_{n+1}, \tau_{n+1}=t_{n+1}$ is often used for the spatio-temporal
argument of the potential. 
These different choices do neither affect the numerical calculation nor the analytical derivation, and different prescriptions can even be translated into each other. 
However, it turns out that the mid-point prescription always
yields the simplest analytic results, and, therefore, we will use it throughout the present paper.

In the following we will switch to the imaginary-time formalism, which is widely used in numerical simulations \cite{ceperley}, since it mitigates problems which are
related to the oscillatory nature of the integrand in the real-time approach. To obtain the real-time from the imaginary-time amplitudes, one would have to perform an inverse Wick-rotation, 
i.e.~a 
suitable analytic continuation of the numerical results. This might be difficult due to the inherent instability of this procedure with respect to statistical noise, which is always present in numerical 
calculations. However, using the imaginary time is justified by the fact that all current applications of this approach are related to quantum statistical physics, which is 
naturally set up in imaginary time, with the inverse temperature $\beta=1/k_BT$ playing the role of the imaginary time. Also, energy spectra and energy eigenfunctions can be efficiently calculated 
in this formalism \cite{diag1, diag2}. Note that the derived analytic expressions for higher-order propagators can be formally transformed from the imaginary- to the real-time axis.
In the next paper \cite{fcpitdp2}, we will demonstrate by treating several simple examples that such analytic expressions can be successfully used for calculating the real-time evolution in the context of the space-discretized approach \cite{diag1, diag2}.

After Wick rotation to the imaginary time, the transition amplitude is expressed as
\begin{equation}
\label{eq:piit}
A(\mathbf a,t_a; \mathbf b, t_b)=\int_{\mathbf q(t_a)=\mathbf a}^{\mathbf q(t_b)=\mathbf b}  {\cal D} \mathbf q(t) \, e^{-\frac{1}{\hbar} S_E[\mathbf q]}\, ,
\end{equation}
where the action is replaced by its imaginary-time counterpart, the Euclidean action
\begin{equation}
\label{eq:SEdef}
S_{\rm E}[\mathbf q]=\int_{t_a}^{t_b} dt\, \left\{ \frac{1}{2}\dot{\mathbf q}^2(t)+V(\mathbf q(t), t)\right\}\, ,
\end{equation}
which represents the energy of the system. To simplify the notation, we will drop the subscript ${\rm E}$ from now on. 
If we consider in Eq.~(\ref{eq:piit}) only diagonal amplitudes, i.e.~$\mathbf a=\mathbf b$, and integrate 
over $\mathbf a$, we obtain the path-integral expression for the partition function
\begin{equation}
Z(\beta)=\mathrm{Tr} \, \left\{ \hat T \exp\left[ - \frac{1}{\hbar} \int_0^{\hbar \beta} d t\, \hat H(t)\right] \right\}
\end{equation}
by setting $t_a=0$ and $t_b=\hbar\beta$.

\section{Path-Integral Calculation of the Propagator}
\label{sec:pi}
In this paper we develop a method for calculating  short-time transition amplitudes for time-dependent potentials 
to high orders in the propagation time. To this end we follow the approach of Ref.~\cite{balazpre} and generalize
the level $p=1$  discretized transition amplitude from Eq.~(\ref{eq:lowest}) in such a way that the exact transition amplitude is written as
\begin{equation}
\label{eq:Aexacta}
A(\mathbf a, t_a; \mathbf b,t_b)=\frac{1}{(2\pi\varepsilon)^{Pd/2}}\, e^{-S^*(\mathbf x, \boldsymbol{\delta}; \varepsilon, \tau)}\, .\\
\end{equation}
Here we use the convention $\hbar=1$, the ideal discretized effective action reads \cite{pla-euler, pre-ideal}
\begin{equation}
\label{eq:Aexactb}
S^*(\mathbf x, \boldsymbol{\delta}; \varepsilon, \tau)
= \frac{\boldsymbol{\delta}^2}{2\varepsilon}+\varepsilon W(\mathbf x, \boldsymbol{\delta}; \varepsilon, \tau)\, ,
\end{equation}
and $W$ 
represents the ideal effective potential, which ensures the exactness of the above expression. 
The latter depends not only on the coordinate mid-point $\mathbf x=(\mathbf a+\mathbf b)/2$, the discretized velocity  $\boldsymbol{\delta}=\mathbf b-\mathbf a$, and 
the time interval $\varepsilon=t_b-t_a$ as already introduced in the case for time-independent potentials, but also on the time mid-point $\tau=(t_a+t_b)/2$, due to the explicit time dependence of the potential.

We will analytically derive a systematic short-time expansion of the effective potential $W$, which provides an
improved convergence for numerically calculating transition amplitudes and 
partition functions as well as other properties of quantum systems with time-dependent potentials. The expansion of the effective potential $W$ to 
order $\varepsilon^{p-1}$ yields the effective action correct to order $\varepsilon^p$, with errors proportional to $\varepsilon^{p+1}$. Due to the normalization factor in the 
expression (\ref{eq:Aexacta}), the total $\varepsilon$-convergence of the amplitude is given by 
\begin{equation}
\label{eq:Apscaling}
A_p(\mathbf a, t_a; \mathbf b,t_b)=A(\mathbf a, t_a; \mathbf b,t_b)+O(\varepsilon^{p+1-Pd/2})\, .
\end{equation}
This $\varepsilon$-scaling of errors is valid if we are interested in calculating short-time transition amplitudes, which is the main objective of this paper. However, if we use such short-time 
transition amplitudes to calculate long-time transition amplitudes through the time-discretization procedure (\ref{eq:complete}), due to successive integrals the normalization factors will be again added, and we will 
get total errors of the order $N\cdot O(1/N^{p+1})=O(1/N^p)$. Although the corresponding effective actions and resulting discretized short-time transition
amplitudes are designated by the index $p$ according to 
their $N$-scaling behavior, we stress that the scaling with respect to the short  propagation time is always given by Eq.~(\ref{eq:Apscaling}).
 
Before we embark upon developing a systematic recursive approach for analytically calculating higher-order effective actions, we first have to study the 
general structure of the 
effective potential, which turns out to be 
more complex than in the case of time-independent quantum systems. In order to do so, we calculate the short-time expansion of the effective 
potential by using an
{\it ab initio} approach introduced in Ref.~\cite{pla-manybody}. To simplify the calculation, we will restrict ourselves in the present 
section to the single-particle one-dimensional case. 
Based on these results we will develop in Sec.~\ref{sec:recursion} the systematic recursive approach for such simple quantum systems, and then extend and 
generalize this procedure to systems with many degrees of freedom.

Following Ref.~\cite{pla-manybody}, we start with changing the variables via $q(t)=\xi(t)+y(t)$, where $\xi(t)$ is some chosen reference trajectory satisfying the same boundary 
conditions as the path $q(t)$, i.e.~$\xi(t_a)=a$, $\xi(t_b)=b$. 
This implies that the new variable $y(t)$ vanishes at the boundaries, i.e.~$y(t_a)=y(t_b)=0$. We also introduce a new time variable 
$s$ by $t=\tau+s$, in which the boundaries are defined by $s_a=-\varepsilon/2$, $s_b=\varepsilon/2$. In the new variables, the kinetic energy functional has the form
\begin{equation}
\label{eq:Ks}
\int_{t_a}^{t_b} dt\, \frac{1}{2}\left(\frac{dq(t)}{dt}\right)^2
=\int_{-\varepsilon/2}^{\varepsilon/2} ds\, \left\{  \frac{1}{2}\dot{\xi}^2(s)+\frac{1}{2}\dot{y}^2(s) -y(s)\ddot{\xi}(s)\right\}\, ,
\end{equation}
where a dot represents a derivative with respect to the time $s$. With this the transition amplitude reads
\begin{equation}
\label{eq:pis}
A(a,t_a; b, t_b)=e^{-S[\xi](\tau)}\, \int_{y(-\varepsilon/2)=0}^{y(\varepsilon/2)=0}  {\cal D} y(s)
\, e^{ -\int_{-\varepsilon/2}^{\varepsilon/2} ds\, \left\{\frac{1}{2} \dot{y}^2(s)+U_\xi(y(s), s)\right\}}\, ,
\end{equation}
where the quantity $U_\xi$ is defined as
\begin{equation}
\label{eq:Udef}
U_\xi(y(s), s)=V(\xi(s)+y(s), \tau+s)-V(\xi(s), \tau)-y(s)\ddot{\xi}(s)\, ,
\end{equation}
and the Euclidean action $S[\xi](\tau)$ for the reference 
trajectory $\xi(s)$ is defined by an expression similar to Eq.~(\ref{eq:SEdef}), but with the time argument 
of the potential being fixed now by the mid-point $\tau$:
\begin{equation}
\label{eq:Sxi}
S[\xi](\tau)=\int_{-\varepsilon/2}^{\varepsilon/2} ds\, \left\{ \frac{1}{2}\dot{\xi}^2(s)+V(\xi(s), \tau)\right\}\, .
\end{equation}
Thus, the transition amplitude reads
\begin{equation}
\label{eq:fpev}
A(a,t_a;b, t_b)=\frac{e^{-S[\xi](\tau)}}{\sqrt{2\pi\varepsilon}}\, \left\langle e^{ -\int_{-\varepsilon/2}^{\varepsilon/2} ds\, U_\xi(y(s), s)}\right\rangle\, ,
\end{equation}
where the path-integral expectation value is defined with respect to the free-particle action:
\begin{equation}
\left\langle \,\bullet \,\right\rangle = \sqrt{2\pi\varepsilon}\int_{y(-\varepsilon/2)=0}^{y(\varepsilon/2)=0}  {\cal D} y(s)\,\bullet\,
e^{ -\int_{-\varepsilon/2}^{\varepsilon/2} ds\, \frac{1}{2} \dot{y}^2(s)} \,.
\end{equation}
The transition amplitude (\ref{eq:fpev}) can then be obtained through a standard calculation of the free-particle 
expectation value by using the Taylor expansion
\begin{eqnarray}
\left\langle e^{ -\int_{-\varepsilon/2}^{\varepsilon/2} ds\, U_\xi(y(s), s)}\right\rangle &=& 1-\int_{-\varepsilon/2}^{\varepsilon/2} ds\, \big\langle U_\xi(y(s), s)\big\rangle\nonumber\\
&&\hspace*{2.5mm}+\quad\frac{1}{2} \int_{-\varepsilon/2}^{\varepsilon/2} ds \int_{-\varepsilon/2}^{\varepsilon/2} ds'\, \big\langle U_\xi(y(s), s)\, U_\xi(y(s'), s')\big\rangle+\ldots \,.
\label{eq:eUexp}
\end{eqnarray}

As we see from Eq.~(\ref{eq:Udef}), the quantity $U_\xi$ has the simplest form if we choose the reference trajectory $\xi$ in such a way that its second derivative 
with respect to the time $s$ vanishes. The 
natural choice is thus the linear trajectory
$$\xi(s)=x+s \delta/ \varepsilon\, ,$$
centered around the mid-point $x=(a+b)/2$. In order to calculate the expectation values in 
Eq.~(\ref{eq:eUexp}), we further have to expand 
\begin{equation}
U_\xi(y(s), s)=V(\xi(s)+y(s), \tau+s)-V(\xi(s), \tau)
\end{equation}
around this reference trajectory according to
\begin{equation}
\label{eq:Uexp}
U_\xi(y(s), s) = \sum_{\genfrac{}{}{0pt}{}{n,m=0}{n+m>0}}^\infty\frac{1}{n!\, m!}\stackrel{(m)}{V}{\hspace*{-1mm}}^{(n)}(\xi(s), \tau)\, y^n(s)\, s^m\, ,
\end{equation}
where, in order to simplify the notation, $(m)$ denotes the order of the partial derivative with respect to the time $s$, and $(n)$ denotes the order of the partial derivative with respect to the spatial coordinate. For a free-particle theory, expectation values of 
the type $\big\langle y^{n_1}(s_1)\, y^{n_2}(s_2)\ldots\big\rangle$
can be easily calculated using the standard generating functional approach. The expectation value $\langle y(s)\rangle$ vanishes due to the symmetry, while the correlator 
%
%\begin{equation}
$\Delta(s, s')=\big\langle y(s)\, y(s')\big\rangle$
%\end{equation}
%
is given by the expression
\begin{equation}
\label{eq:prop}
\Delta(s, s')=\frac{\theta(s-s')}{\varepsilon}\left(\frac{\varepsilon}{2}-s\right)\left(\frac{\varepsilon}{2}+s'\right) + (s\leftrightarrow s')\, ,
\end{equation}
and higher expectation values can be found in Ref.~\cite{pla-manybody}.

In order to obtain an expansion of the transition amplitude in the propagation time
$\varepsilon$, one has to consider Eq.~(\ref{eq:eUexp}) and to take into account the powers 
of $\varepsilon$ of all terms to identify the relevant terms at the desired level $p$. To illustrate this, let us look at the term linear in $U_\xi$
\begin{equation}
\label{eq:Ulin}
\int_{-\varepsilon/2}^{\varepsilon/2} ds\, \big\langle U_\xi(y(s), s)\big\rangle
= \int_{-\varepsilon/2}^{\varepsilon/2} ds\, \sum_{\genfrac{}{}{0pt}{}{n,m=0}{n+m>0}}^\infty\frac{1}{n!\, m!}\stackrel{(m)}{V}{\hspace*{-1mm}}^{(n)}(\xi(s), \tau)\, 
\big\langle y^n(s)\big\rangle \, s^m\, ,
\end{equation}
where the 
integration over $s$ yields terms proportional to $\varepsilon^{m+1}$ times the contribution of the expectation value term. For example, from Eq.~(\ref{eq:prop}) we see that 
the $n=2$ expectation value would yield an additional $\varepsilon$ power, amounting to a total $\varepsilon^{m+2}$ dependence of the term corresponding to a 
given $m$ and $n=2$. Since we are interested to calculate all terms to a given order in $\varepsilon$, we have to 
carefully inspect all possible terms and to select the appropriate ones at a given level $p$. As an example we write down all terms which contribute to the expectation value 
(\ref{eq:eUexp}) to order $O(\varepsilon^4)$:

\begin{eqnarray}
&&\hspace*{-1.5cm}
\left\langle e^{ -\int_{-\varepsilon/2}^{\varepsilon/2} ds\, U_\xi(y(s), s)}\right\rangle =
1 -\int_{-\varepsilon/2}^{\varepsilon/2} ds\, \left\{\frac{1}{2}V''(\xi(s), \tau) \big\langle y^2(s)\big\rangle +
\frac{1}{24}V^{(4)}(\xi(s), \tau) \big\langle y^4(s)\big\rangle \right.\nonumber\\
&&\hspace*{15mm}\left. +\quad \dot{V}(\xi(s), \tau) s+\frac{1}{2}\dot{V}^{(2)}(\xi(s), \tau) \big\langle y^2(s)\big\rangle s
+\frac{1}{2}\ddot{V} (\xi(s), \tau) s^2\right\}\nonumber\\
&&\hspace*{15mm}+\quad \frac{1}{2}\int_{-\varepsilon/2}^{\varepsilon/2} ds \int_{-\varepsilon/2}^{\varepsilon/2} ds'\, V'(\xi(s), \tau) V'(\xi(s'), \tau)\, \big\langle y(s)y(s')\big\rangle +O(\varepsilon^4)\, .
\label{eq:O4exp}
\end{eqnarray}
In the above expression, the correlators $\langle y^2(s)\rangle=\Delta(s, s)$ and $\langle y(s) y(s')\rangle=\Delta(s, s')$ are given by Eq.~(\ref{eq:prop}), and the expectation 
value $\langle y^4(s)\rangle$ can be directly determined using either the generating functional method or the Wick rule, 
yielding $3\Delta^2(s, s)$. In order to be able to calculate the remaining integrals over 
$s$ and $s'$, we need to expand also 
the potential $V$ and its derivatives with respect to the first argument $\xi(s)=x+s \delta/ \varepsilon$ around the mid-point $x$. The required number of 
terms in this expansion, contributing to the desired order of $\varepsilon$, is obtained by taking into account the diffusion relation $\delta^2\sim\varepsilon$, which is valid 
for small propagation times $\varepsilon$ and has been demonstrated to yield a consistent expansion of short-time transition
amplitudes \cite{balazpre}. The expansion of the potential into 
power series in $s \delta / \varepsilon$ gives an additional polynomial $s$-dependence, which finally
allows an analytic calculation of all integrals in Eq.~(\ref{eq:O4exp}). When this 
is done, we obtain the following expression for the expectation value (\ref{eq:eUexp})
\begin{eqnarray}
\label{Resu}
\left\langle e^{ -\int_{-\varepsilon/2}^{\varepsilon/2} ds\, U_\xi(y(s), s)}\right\rangle &=&
1-V''\frac{\varepsilon^2}{12} - \dot{V}'\frac{\delta\varepsilon^2}{12}
-V^{(4)}\frac{\varepsilon^3}{240}-\ddot{V}\frac{\varepsilon^3}{24}
+V'^2\frac{\varepsilon^3}{24}-V^{(4)}\frac{\delta^2\varepsilon^2}{480}\nonumber\\
&&\hspace*{2.5mm}-\quad \dot{V}^{(3)}\frac{\delta\varepsilon^3}{240}- \dot{V}^{(3)}\frac{\delta^3\varepsilon^2}{480}+ O(\varepsilon^4)\, ,
\end{eqnarray}
where the potential $V$ as well as its spatial and temporal derivatives are evaluated at the 
mid-point $x, \tau$, i.e.~$V=V(x, \tau)$. Note that we have retained in Eq.~(\ref{Resu}) only those terms whose  
order is less than $\varepsilon^4$. Taking into account the diffusion
relation $\delta^2\sim\varepsilon$, the terms proportional to $\varepsilon^3$ and $\delta^2\varepsilon^2$ are considered 
to be of the same order.

Combining Eqs.~(\ref{eq:Aexacta}) and (\ref{eq:fpev}), we see that the ideal effective action can be expressed by
its short-time expression (\ref{eq:Sxi}) and the expectation value (\ref{eq:eUexp}) according to
\begin{equation}
\label{eq:SstarS}
S^*(x, \delta; \varepsilon, \tau)=S[\xi](\tau)-\log\left\langle e^{ -\int_{-\varepsilon/2}^{\varepsilon/2} ds\, U_\xi(y(s), s)}\right\rangle\, .
\end{equation}
After expanding the potential $V$ around the mid-point $x$ in the action $S[\xi](\tau)$ in Eq.~(\ref{eq:Sxi}), we obtain its short-time expansion:
\begin{equation}
\label{eq:Sxiexp}
S[\xi](\tau)=\frac{\delta^2}{2\varepsilon}+V\varepsilon+V''\frac{\delta^2\varepsilon}{24}+V^{(4)}\frac{\delta^4\varepsilon}{1920}+O(\varepsilon^4)\, .
\end{equation}
This allows us to calculate the short-time expansion of the effective potential $W$ from Eqs.~(\ref{eq:Aexactb}) and (\ref{Resu})--(\ref{eq:Sxiexp}). 
For example, up to level $p=3$ we get
\begin{equation}
W_3 (x, \delta; \varepsilon, \tau) =
V+V''\frac{\varepsilon}{12} +V''\frac{\delta^2}{24}+ \dot{V}'\frac{\delta\varepsilon}{12}\label{eq:Wp3}
+V^{(4)}\frac{\varepsilon^2}{240}+\ddot{V}\frac{\varepsilon^2}{24}-V'^2\frac{\varepsilon^2}{24}+V^{(4)}\frac{\delta^2\varepsilon}{480}+V^{(4)}\frac{\delta^4}{1920}\, .
\end{equation}
Numerically, such a result allows to a speed up the calculation of transition amplitudes, since the errors can be substantially reduced using analytic expressions for higher level effective actions.

Compared to our previous results for effective actions of time-independent potentials $V$ in Ref.~\cite{balazpre}, 
we see that new terms appear which contain time derivatives of the potential. 
In particular, we observe the emergence of terms with odd powers of the discretized velocity $\delta$, which was previously not the case. In fact, for time-independent 
potentials we 
have shown in Ref.~\cite{balazpre} 
that the effective potential is symmetric in $\delta$, which leaves only even powers of $\delta$ in its short-time expansion. Here, however, also 
odd powers of $\delta$ survive due to the explicit time dependence of the potential. We also recognize that all new terms are proportional to time derivatives of the potential and 
vanish in the time-independent case, thus reducing the effective action to the previous expressions in Ref.~\cite{balazpre}.

Therefore, the correct systematic short-time expansion of the effective potential turns out to have the form
\begin{equation}
\label{eq:Wdouble}
W(x,\delta;\varepsilon, \tau)=\sum_{m=0}^{\infty}\sum_{k=0}^{m}\left\{c_{m,k}(x, \tau)\,\varepsilon^{m-k}\left(\frac{\delta}{2}\right)^{2k}
+ c_{m+1/2,k}(x, \tau)\,\varepsilon^{m-k}\left(\frac{\delta}{2}\right)^{2k+1}\right\} \, .
\end{equation}
Here $\delta/2$ is used as the expansion parameter in order to have expansion coefficients $c$ which are
defined consistently with Ref.~\cite{balazpre}. Such an expansion allows that the level $p$ effective 
action is written as the sum of terms corresponding to $m=0,1,\ldots, p-1$. Note that for $m=p-1$ we need to take into account only the even-power terms 
$c_{p-1, k}(x, \tau)\,\varepsilon^{p-1-k}(\delta/2)^{2k}$. The odd-power terms $c_{p-1/2,k}(x, \tau)\,\varepsilon^{p-1-k}(\delta/2)^{2k+1}$ are proportional to $\varepsilon^{p-1/2}$, 
i.e.~they are of  higher order than the required $\varepsilon^{p-1}$ for level $p$ effective action. For this reason, the correct expansion of the effective potential at  
level $p$ is given by
\begin{equation}
\label{eq:Wpdouble}
\hspace*{-10mm}
W_p(x,\delta;\varepsilon, \tau)=\sum_{m=0}^{p-1}\sum_{k=0}^{m}c_{m,k}(x, \tau)\,\varepsilon^{m-k}\left(\frac{\delta}{2}\right)^{2k}
+ \sum_{m=0}^{p-2}\sum_{k=0}^{m} c_{m+1/2,k}(x, \tau)\,\varepsilon^{m-k}\left(\frac{\delta}{2}\right)^{2k+1},
\end{equation}
and it provides the convergence of transition amplitudes according to Eq.~(\ref{eq:Apscaling}).

\section{Forward and Backward Schr\"odinger Equation}
\label{sec:sch}
In this and in the next section we will use the latter result (\ref{eq:Wdouble}) to develop a systematic recursive approach for analytically calculating effective actions to high levels $p$ for time-dependent potentials. To this end, in this section we rederive both the forward and the backward Schr\"odinger equation for the transition amplitude and then use them to derive corresponding differential equations  for the effective potential $W$. Afterwards, in Sec.~\ref{sec:recursion} we use Eq.~(\ref{eq:Wdouble}) to solve the derived equations for $W$ and to obtain recursion relations for single-particle one-dimensional systems.

The evolution operator for quantum system in a time-dependent potential is given by Eq.~(\ref{eq:UT}), or, in imaginary time,
\begin{equation}
\label{eq:UTi}
\hat U(t_a\to t_b)=\hat{T}\exp\left\{-\int_{t_a}^{t_b}dt\, \hat H(\hat{\mathbf p}, \hat{\mathbf q}, t) \right\}\, .
\end{equation}
Thus, the evolution operator obeys the differential equation
\begin{equation}
\label{eq:dtbU}
\partial_{t_b}\, \hat U(t_a\to t_b)=- \hat H(\hat{\mathbf p}, \hat{\mathbf q}, t_b)\, \hat U(t_a\to t_b)\, ,
\end{equation}
and, similarly,
\begin{equation}
\label{eq:dtaU}
\partial_{t_a}\, \hat U(t_a\to t_b)=\hat U(t_a\to t_b)\, \hat H(\hat{\mathbf p}, \hat{\mathbf q}, t_a)\, .
\end{equation}
If we determine from Eq.~(\ref{eq:dtbU}) the
matrix elements which correspond to the transition amplitude Eq.~(\ref{eq:AU}), we obtain the forward Schr\"odinger equation for the transition 
amplitude
\begin{equation}
\label{eq:dtbA}
\partial_{t_b}\, A(\mathbf a,t_a; \mathbf b, t_b) =- \hat H_b\, A(\mathbf a,t_a; \mathbf b, t_b)\, ,
\end{equation}
where $\hat H_b$ stands for the coordinate-space Hamilton operator $\hat H_b=H(-i \boldsymbol{\partial}_{\mathbf b}, \mathbf b, t_b)$, in which momentum and position operators 
are replaced by their coordinate-space representations at $\mathbf b$. To obtain the analogous equation for the derivative with respect to the initial time $t_a$, we have to 
take into account that the imaginary-time transition amplitudes as well as its time derivative are real. With this Eq.~(\ref{eq:dtaU}) yields at first
\begin{equation}
\langle\mathbf b|\partial_{t_a}\, \hat U(t_a\to t_b) |\mathbf a\rangle
=H(-i \boldsymbol{\partial}_{\mathbf a}, \mathbf a, t_a)\,  \langle\mathbf a|\hat U^\dagger(t_a\to t_b) |\mathbf b\rangle\, .
\end{equation}
Since we have in addition
\begin{equation}
\langle\mathbf a|\hat U^\dagger(t_a\to t_b) |\mathbf b\rangle=A(\mathbf a,t_a; \mathbf b, t_b)\, ,
\end{equation}
we finally get the backward  Schr\"odinger equation for the transition  amplitude
\begin{equation}
\label{eq:dtaA}
\partial_{t_a}\, A(\mathbf a,t_a; \mathbf b, t_b) = \hat H_a\, A(\mathbf a,t_a; \mathbf b, t_b)\, ,
\end{equation}
where $\hat H_a=H(-i \boldsymbol{\partial}_{\mathbf a}, \mathbf a, t_a)$ is defined analogously as $\hat H_b$.

In the next step we change the original time variables $t_a$ and $t_b$ to the mid-point $\tau$ and the propagation time $\varepsilon$, which converts
(\ref{eq:dtbA}) and (\ref{eq:dtaA}) to
\begin{eqnarray}
\label{eq:depsA}
\left[\partial_\varepsilon +\frac{1}{2}(\hat H_a + \hat H_b)\right] A(\mathbf a,t_a; \mathbf b, t_b)&=&0\, ,\\
\label{eq:dtauA}
\left[\partial_\tau +(\hat H_b - \hat H_a)\right] A(\mathbf a,t_a; \mathbf b, t_b)&=&0\, .
\end{eqnarray}
Subsequently, we also change the spatial variables $\mathbf a$ and $\mathbf b$ to the mid-point $\mathbf x$ and the
discretized velocity $\bar{\mathbf x}=\boldsymbol{\delta}/2$, thus Eqs.~(\ref{eq:depsA}) and (\ref{eq:dtauA}) read then
\begin{eqnarray}
\label{eq:depsxbarxA}
\left[\partial_\varepsilon-\frac{1}{8}\,\partial^2-\frac{1}{8}\,\bar\partial^2
+\frac{1}{2}\, (V_++V_-)\right]A(\mathbf x, \bar{\mathbf x}; \varepsilon, \tau)&=&0\, ,\\
\label{eq:dt}
\left[\partial_\tau -\frac{1}{2}\partial\bar\partial+V_+ -V_- \right]A(\mathbf x, \bar{\mathbf x}; \varepsilon, \tau) &=& 0\, ,
\end{eqnarray}
where we have introduced $V_\pm=V\left(\mathbf x\pm\bar{\mathbf x}, \tau\pm\frac{\varepsilon}{2}\right)$ as abbreviations, and the $P\times d$-dimensional 
Laplacians $\boldsymbol{\partial}_{\mathbf x}\cdot \boldsymbol{\partial}_{\mathbf x}=\partial^2$ over coordinates $\mathbf x$ and $\boldsymbol{\partial}_{\bar{\mathbf x}} 
\cdot \boldsymbol{\partial}_{\bar{\mathbf x}} =\bar\partial^2$ over coordinates $\bar{\mathbf x}$,
as well as the mixed Laplacian $\boldsymbol{\partial}_{\mathbf x}\cdot \boldsymbol{\partial}_{\bar{\mathbf x}}=\partial\bar\partial$. 
Note that we 
will not use Eq.~(\ref{eq:dt}) in our further calculation of the short-time transition amplitude. This equation describes the dynamics caused only by the presence of the 
explicit time-dependence of the potential, as is indicated by the derivative with respect to the time mid-point $\tau$.

If we now express the transition amplitude (\ref{eq:Aexacta}) by using the effective potential in (\ref{eq:Aexactb}) and the new spatio-temporal variables, we get
\begin{equation}
\label{eq:AWxbarx}
A(\mathbf x, \bar{\mathbf x}; \varepsilon, \tau)=
\frac{1}{(2\pi\varepsilon)^{Pd/2}}e^{-\frac{2}{\varepsilon}\bar{\mathbf x}^2-\varepsilon W(\mathbf x, \bar{\mathbf x}; \varepsilon, \tau)}\, .
\end{equation}
Substituting this expression into the differential equation for the transition amplitude (\ref{eq:depsxbarxA}), we obtain a corresponding 
differential equation for the effective potential:
\begin{equation}
W+\bar{\mathbf x}\cdot\bar{\boldsymbol{\partial}}\,W+\varepsilon\partial_\varepsilon W
-\frac{1}{8}\,\varepsilon\partial^2 W-\frac{1}{8}\,\varepsilon\bar\partial^2 W
+\frac{1}{8}\,\varepsilon^2(\boldsymbol{\partial} W)^2
+\frac{1}{8}\,\varepsilon^2(\bar{\boldsymbol{\partial}} W)^2=\frac{1}{2}\, (V_++V_-)\, .
\label{eq:Weq}
\end{equation}

This equation is formally identical to the corresponding equation (29) from Ref.~\cite{balazpre}, and in the limit of the time-independent potential we recover the previously 
derived result. Eq.~(\ref{eq:Weq}) can now be used to develop a recursive approach for calculating the effective potential $W$ by using the double power 
series (\ref{eq:Wdouble}) in both $\varepsilon$ and $\bar{\mathbf x}$.

\section{Recursive  Calculation of the Propagator for One-dimen\-sional Systems}
\label{sec:recursion}
In this section we consider a single-particle one-dimensional system in a given external time-dependent potential $V$, and fully develop recursive approach for calculation of the short-time expansion of the effective potential $W$. In this case, the right-hand side of Eq.~(\ref{eq:Weq}) can be expanded in a double power series of the form
\begin{equation}
\frac{1}{2}\, (V_++V_-)=\frac{1}{2}\sum_{k, m\geq 0}^\infty\frac{1}{k!\, m!}
\stackrel{(m)}{V}{\hspace*{-1mm}}^{(k)}\, \left(\frac{\varepsilon}{2}\right)^m \bar x^k \Big\{1+(-1)^{m+k}\Big\}\, .
\label{double}
\end{equation}
We see that $m$ and $k$ must be of the same parity in order to yield a non-zero contribution. Therefore, we introduce the quantity $\Pi(m, k)$, which is equal to one if $m-k$ is 
even, and vanishes otherwise. This allows us to reorganize terms of the double power series (\ref{double}) in a form which corresponds to the expansion of the effective potential
in Eq.~(\ref{eq:Wdouble}):
\begin{eqnarray}
&&\hspace*{-12mm}\frac{1}{2}\, (V_++V_-)=\nonumber\\
&&\hspace*{-4mm}\sum_{m=0}^\infty\sum_{k=0}^m\left\{ \frac{\Pi(m, k)\, \varepsilon^{m-k}\, \bar x^{2k}}{(2k)!\, (m-k)!\, 2^{m-k}}\stackrel{(m-k)}{V}{\hspace*{-1mm}}^{(2k)}
+ \frac{[1-\Pi(m, k)]\,  \varepsilon^{m-k}\, \bar x^{2k+1}}{(2k+1)!\, (m-k)!\, 2^{m-k}}\stackrel{(m-k)}{V}{\hspace*{-1mm}}^{(2k+1)} \right\}\, .
\label{eq:Vpmexp}
\end{eqnarray}
In order to solve Eq.~(\ref{eq:Weq}) with (\ref{eq:Vpmexp}), 
we now substitute
the effective potential $W$ with its double expansion (\ref{eq:Wdouble}). By comparing terms with even and odd powers of $\bar x$, we
obtain algebraic equations which determine the coefficients $c_{m, k}$ and $c_{m+1/2, k}$. After a straight-forward calculation we get
\begin{eqnarray}
&&\hspace*{-14mm}
8(m+k+1)\, c_{m,k} = 8 \frac{\Pi(m, k) \stackrel{(m-k)}{V}{\hspace*{-1mm}}^{(2k)}}{(2k)!\, (m-k)!\, 2^{m-k}} + (2k+2) (2 k+1)\, c_{m,k+1}+c_{m-1,k}''\nonumber\\
&&\hspace*{-7mm}-\sum_{l, r}\Big\{c_{l,r}'\,c_{m-l-2,k-r}'+c_{l+1/2,r}'\,c_{m-l-5/2,k-r-1}'+ 2r\, (2k-2r+2)\, c_{l,r}\,c_{m-l-1,k-r+1} \nonumber\\
&&\hspace*{-7mm}+ (2r+1)\, (2k-2r+1)\, c_{l+1/2,r}\,c_{m-l-3/2,k-r}\Big\}\, ,
\label{eq:1dreceven}\\
&&\hspace*{-14mm}
8(m+k+2)\, c_{m+1/2,k} = 8 \frac{[1-\Pi(m, k)] \stackrel{(m-k)}{V}{\hspace*{-1mm}}^{(2k+1)}}{(2k+1)!\, (m-k)!\, 2^{m-k}}+ (2k+3) (2k+2)\, c_{m+1/2,k+1}+c_{m-1/2,k}''
\nonumber\\
&&\hspace*{-7mm}-\sum_{l, r}\Big\{c_{l,r}'\,c_{m-l-3/2,k-r}' 
+c_{l+1/2,r}'\,c_{m-l-2,k-r}' +(2r+1)\, (2k-2r+2)\, c_{l+1/2,r}\,c_{m-l-1,k-r+1}\nonumber\\
&&\hspace*{-7mm}+ 2r\, (2k-2r+3)\, c_{l,r}\,c_{m-l-1/2,k-r+1}\Big\}\, .
\label{eq:1drecodd}
\end{eqnarray}

In the above relations, sums over $l$ and $r$ are restricted by the non-negativity of indices for both even and odd-power coefficients $c$. It is immediately seen 
in the case of a time-independent potential that the 
recursion for even-power coefficients $c_{m, k}$ in Eq.~(\ref{eq:1dreceven}) reduces to the previously derived recursion of Ref.~\cite{balazpre}, 
while the odd-power recursion (\ref{eq:1drecodd})
automatically renders all coefficients $c_{m+1/2, k}$ to be zero. 

In order to efficiently solve these algebraic equations at a given level $p$, we recall the corresponding hierarchical level $p$ expansion (\ref{eq:Wpdouble}). In such an expansion, we 
look only for even-power coefficients $c_{m, k}$ with $0\leq m\leq p-1$, and odd-power coefficients $c_{m+1/2, k}$ with $0\leq m\leq p-2$, while 
the index $k$ is always restricted by $0\leq k\leq m$. To simplify the calculation, we formally set all the coefficients with higher values of indices to zero. As in case 
for a time-independent potential in Ref.~\cite{balazpre}, we are then able to calculate explicitly diagonal even-order coefficient $c_{m, m}$  for any $m$:
\begin{equation}
\label{eq:cmm}
c_{m,m}=\frac{V^{(2m)}}{(2m+1)!}\, .
\end{equation}
Correspondingly, we obtain for the diagonal odd-power coefficients the general result
\begin{equation}
\label{eq:cmmq}
c_{m+1/2, m}=0\, .
\end{equation}
Thus, the recursion relations (\ref{eq:1dreceven}) and (\ref{eq:1drecodd}) represent together with (\ref{eq:cmm}) and (\ref{eq:cmmq}) a
closed set of algebraic equations, which completely determines the 
coefficients $c$ for the effective potential at a given level $p$.

To illustrate how the recursive procedure works in detail, we solve the single-particle recursive relations up
to order $p=3$ and demonstrate that we obtain the same result (\ref{eq:Wp3}) as in the previous section, where we used the 
path-integral approach to calculate the effective potential. For the level $p=3$ we need coefficients $c$ with the first index $m$ in the range $0, \ldots ,p-1$, i.e.~$m=0, 1, 2$.
For a given $m\leq p-1$ we start with calculating even-power coefficients $c_{m, k}$ by using the recursive relation (\ref{eq:1dreceven}) in the order 
$k=m,m-1,\ldots ,0$, where we take into account (\ref{eq:cmm}). 
Then we proceed with the calculation of odd-power coefficients $c_{m+1/2, k}$ by using Eq.~(\ref{eq:1drecodd}) for the order $k=m-1,\ldots ,0$ and by taking 
into account (\ref{eq:cmmq}). We also recall that odd-power coefficients need not be calculated for $m=p-1$, i.e.~in the last step of the recursive procedure for $m=p-1$ 
we only calculate even-power coefficients. Following this procedure, we immediately from get (\ref{eq:cmm}) for $m=0$
\begin{equation}
\label{eq:c00}
c_{0,0} = V \, .
\end{equation}
The coefficient $c_{1/2, 0}=0$ follows trivial from (\ref{eq:cmmq}). For $m=1$ we get the following even-power coefficients, corresponding to the values $k=1$ and $k=0$:
\begin{eqnarray}
\label{eq:c11}
&&c_{1,1} = \frac{1}{6}V'' \, ,\\
\label{eq:c10}
&&c_{1,0} = \frac{1}{16}c''_{0,0} + \frac{1}{8}c_{1,1} =  \frac{1}{12}V''\, .
\end{eqnarray}
For the only non-trivial $m=1$ odd-power coefficient $c_{3/2, 0}$ we get
\begin{equation}
\label{eq:c320}
c_{3/2,0} =\frac{1}{4}c_{3/2, 1}+\frac{1}{24}c''_{1/2, 0}+\frac{1}{6}\dot{V}'=\frac{1}{6}\dot{V}'\, .
\end{equation}
Similarly, we find the $m=2$ even-power coefficients by using Eqs.~(\ref{eq:1dreceven}) and (\ref{eq:cmm}):
\begin{eqnarray}
&&\hspace*{-1cm}
c_{2,2} = \frac{1}{120}V^{(4)}\, ,\\
&&\hspace*{-1cm}
c_{2,1} = \frac{1}{32}c''_{1,1}+\frac{3}{8}c_{2,2}=\frac{1}{120}V^{(4)}\, ,\\
&&\hspace*{-1cm}
\label{eq:c20}
c_{2,0} =\frac{1}{24}\ddot{V}-\frac{1}{24}c'^2_{0,0}+  \frac{1}{24}c''_{1,0}+\frac{1}{12}c_{2,1}=\frac{1}{24}\ddot{V} + \frac{V^{(4)}}{240}-\frac{V'^2}{24}\, .
\end{eqnarray}
This result is already sufficient to construct the effective potential $W_{p=3}$. If we insert the calculated coefficients in the expansion (\ref{eq:Wpdouble}) at the given level $p=3$, 
we reobtain the same expression as in Eq.~(\ref{eq:Wp3}). Comparing the calculated even-power coefficients $c_{m, k}$ with the previously calculated ones for the 
time-independent potential in Ref.~\cite{balazpre}, we see 
that they coincide if we set all time derivatives of the potential to zero, as expected, and that all odd-power coefficients vanish.

We stress that the recursive approach is far more efficient than the path-integral calculation presented in the previous section. For example, if one wants to extend a 
level $p$ calculation to a higher order $p'=p+1$, this requires in the path-integral approach not only to take into account the next term in the expansion  
(\ref{eq:O4exp}), but also each previously calculated expectation value term has to be redone to one order in $\varepsilon$ higher. The complexity of this algorithm prevents 
its efficient implementation. However, in the present recursive approach, all we need to do is to calculate one additional order of odd-power coefficients $c_{p-1/2, k}$, which 
corresponds to $m=p'-2=p-1$, and even-power coefficients $c_{p, k}$, which corresponds to $m=p'-1=p$. 
To do this, we just apply the recursive relations and use previously calculated lower-order 
coefficients $c$. For instance, in order to obtain level $p'=4$, we proceed first with calculating $m=p'-2=2$ odd-power coefficients $c_{5/2, k}$. 
The highest coefficient $c_{5/2, 2}=0$ is automatically equal to 
zero due to (\ref{eq:cmmq}), while for lower-level coefficients we get
\begin{eqnarray}
&&\hspace*{-12mm}
c_{5/2,1} = \frac{1}{2}c_{5/2,2}+\frac{1}{40}c''_{3/2,1}+\frac{1}{60}\dot{V}^{(3)}=\frac{1}{60}\dot{V}^{(3)}\, ,\\
&&\hspace*{-12mm}
c_{5/2,0} =\frac{3}{16}c_{5/2, 1}+ \frac{1}{32}c''_{3/2,0}-\frac{1}{16}c'_{1/2,0}c'_{0, 0}-\frac{1}{8}c_{1/2,0}c_{1, 1}=\frac{1}{120}\dot{V}^{(3)}\, .
\end{eqnarray}
To completely determine the $p'=4$ effective action, we then calculate even-power $m=p'-1=3$ coefficients $c_{3, k}$, and obtain from Eqs.~(\ref{eq:1dreceven}) and (\ref{eq:cmm})
\begin{eqnarray}
&&\hspace*{-9mm}
c_{3,3} = \frac{1}{5040}V^{(6)} \, ,\\
&&\hspace*{-9mm}
c_{3,2} =  \frac{1}{3360}V^{(6)} \, ,\\
&&\hspace*{-9mm}
c_{3,1} =  \frac{1}{3360}V^{(6)} +   \frac{1}{80}\ddot{V}''  -  \frac{1}{360} V''^2 -  \frac{1}{120}V' V^{(3)} ,\\
&&\hspace*{-9mm}
c_{3,0} = \frac{1}{6720}V^{(6)} +   \frac{1}{480}\ddot{V}''  -  \frac{1}{360} V''^2 -  \frac{1}{120}V' V^{(3)} .
\label{c30}
\end{eqnarray}

The outlined procedure continues in the same way for higher levels $p$. We have automatized this procedure and implemented it in our SPEEDUP code \cite{speedup} using the Mathematica 7.0 package 
\cite{mathematica} for symbolic calculus. Using this we have determined the effective action for a one-dimensional particle in a general time-dependent potential up to the level $p=20$. Such 
calculation requires around 2 GB of memory and approximately 1.5 hours of CPU time on 2.33 GHz Intel Xeon E5345 processor.
Although the effective actions grow in complexity with level $p$, the Schr\" odinger equation method for calculating the discrete-time effective actions turns out to be 
extremely efficient. The value of $p=20$ far surpasses the previously obtained best result known in literature 
of $p=6$ \cite{bayepre}, and is limited practically only by the sheer size of the 
expression for the effective action of a general theory at such a high level. The whole technique can be pushed even much further when working on specific potential classes
as, for instance, polynomial potentials, where 
higher spatial and temporal derivatives have a simple form. However, if this is not the case, the obtained general expressions for the effective 
potential can be used for any given potential.

In practical applications, the increasing complexity of expressions for higher-level effective actions leads to a
corresponding increase in the necessary computation time. For levels $p\lesssim 10$ 
\cite{prl-speedup, prb-speedup}, the increase in the computation time is minimal compared to the obtained decrease in the error of numerically calculated transition amplitudes, thus it can be used to 
speed up the calculation at a given level of accuracy. However, for higher values of $p$ the error may be more efficiently reduced by decreasing the time step, i.e.~by using a larger number of 
Monte Carlo discretization steps. Since the complexity of effective actions strongly depends on the concrete form of the given potential, 
the optimal values of both the level $p$ and the size of the time step have to be
estimated from a series of scaled-down numerical simulations. For example, level $p=21$ and a certain time step have turned out to be optimal for
studying fast-rotating Bose-Einstein condensates in an anharmonic trapping potential \cite{becpla}.

We will generalize the presented approach to many-body systems in the next paper \cite{fcpitdp2}, which will also include generalization of the imaginary-time to the real-time formalism, suitable for study of dynamics of quantum systems.

\section{Numerical Verification}
\label{sec:numver}

In order to numerically verify the derived analytical results, we have studied several simple models with time-dependent potentials. We have used the modified version of 
the Monte Carlo SPEEDUP \cite{speedup} code, which implements higher order effective actions in the C programming language. In order to be able to resolve decreasingly small 
errors associated with higher levels $p$ of the effective potential, we decided to consider at first some exactly solvable potentials. 

\begin{figure}[!t]
\centering
\includegraphics[width=8.1cm]{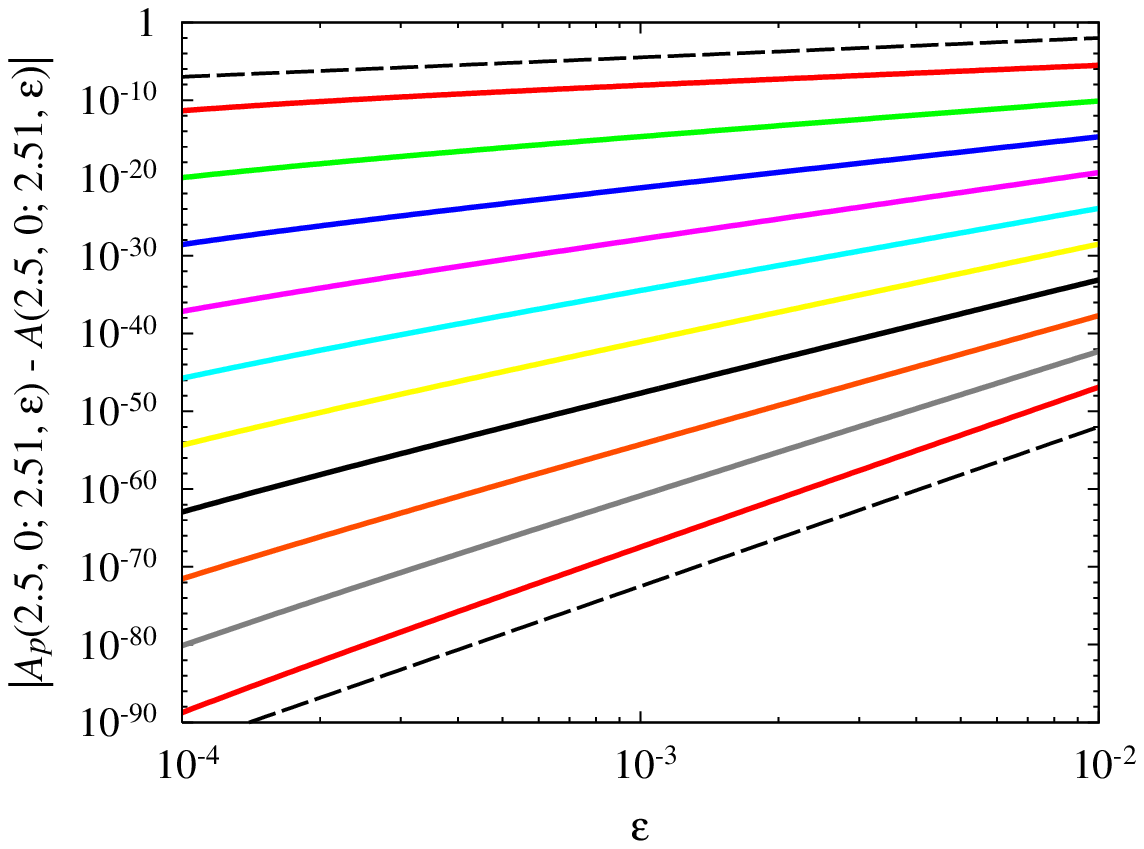}
\includegraphics[width=8.1cm]{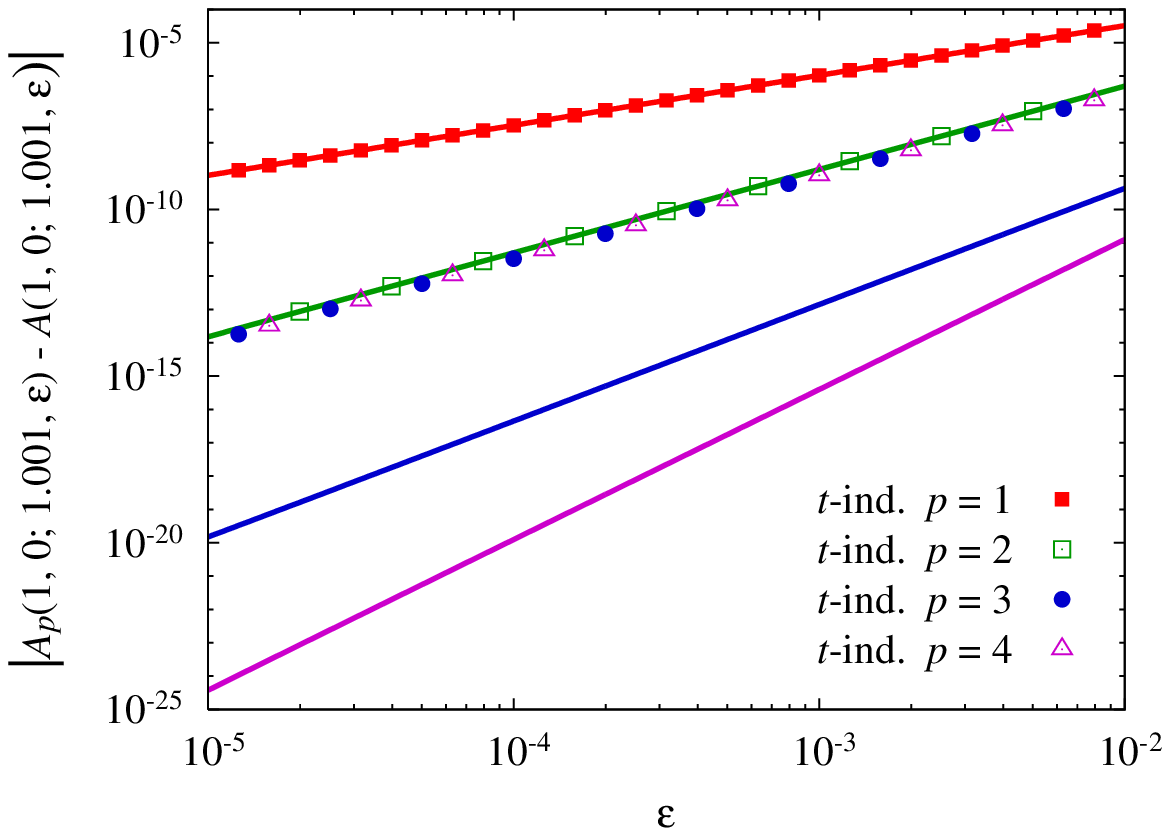}
\caption{Deviations of amplitudes as functions of propagation time $\varepsilon=t_b-t_a$ for the harmonic oscillator 
(\ref{eq:hogrosche}) with $\omega=1$, rescaled with the Grosche factor $\zeta(t)=\sqrt{t^2+1}$. (left) Deviations of amplitudes $|A_p(2.5, 0; 2.51, \varepsilon)-A(2.5, 0; 2.51, \varepsilon)|$ as functions 
of $\varepsilon$, calculated analytically for $p=2, 4, 6, 8, 10, 12, 14, 16, 18, 20$ from top to bottom. The dashed lines are proportional to $\varepsilon^{2.5}$ and $\varepsilon^{20.5}$ and  
demonstrate the perfect scaling of the corresponding level $p=2$ and $p=20$ results. (right) Comparison of deviations $|A_p (1, 0; 1.001, \varepsilon)-A
(1, 0; 1.001, \varepsilon)|$ calculated using the correct level $p=1, 2, 3, 4$ effective potentials (full lines, top to bottom) with the deviations obtained for the same levels $p$ 
of previously derived effective actions \cite{balazpre} for the case of time-independent potential. Deviations for different levels $p$ of time-independent effective actions 
correspond to different point types.}
\label{fig:hogrosche}
\end{figure}

In Ref.~\cite{groschepla} it was shown that a
generalized Duru-Kleinert transformation \cite{kleinertbook,storchak,diploma}
allows to map the transition amplitude for a time-independent potential $V(x)$ to the corresponding transition amplitude for the time-dependent potential
\begin{equation}
\label{eq:grosche}
V_{\rm G}(x, t)=\frac{1}{\zeta^2(t)} V\left(\frac{x}{\zeta(t)}\right)\, ,
\end{equation}
where $\zeta^2(t)=\zeta_0+\zeta_1 t +\zeta_2 t^2$ is a general quadratic polynomial in time. Fig.~\ref{fig:hogrosche} presents the numerical results for the time-dependent harmonic 
oscillator potential, where the time dependence is introduced by using the Grosche rescaling factor $\zeta(t)=\sqrt{1+t^2}$:
\begin{equation}
\label{eq:hogrosche}
V_{\rm G, HO}(x, t)=\frac{\omega^2 x^2}{2 (1+t^2)^2}\, .
\end{equation}

In the left plot of Fig.~\ref{fig:hogrosche} we see that the obtained discretized amplitudes converge to the continuum limit systematically faster and faster when we use higher level effective actions. 
This log-log plot demonstrates that the analytically derived law $\varepsilon^{p+1/2}$ for deviations of single-particle discretized transition
amplitudes in $d=1$ from the continuum transition amplitudes 
holds perfectly, 
which verifies  numerically our analytical results. The deviations are calculated using the analytically known continuous transition amplitude \cite{groschepla}. The graph on the right of 
Fig.~\ref{fig:hogrosche} illustrates the importance of terms with time derivatives of the potential in higher-order effective actions. In this plot we show deviations of discretized transition 
amplitudes calculated using the effective actions for time-independent potentials, derived in Ref.~\cite{balazpre}, and compare them with the deviations of discretized 
transition amplitudes calculated with the 
correct effective actions, derived here for the case of time-dependent potentials. As can be seen from the graph, time-independent effective actions do not improve results after level $p=2$, due to 
the fact that terms, containing time derivatives of the potential, are not systematically eliminated from deviations when such effective actions are used. If we use expressions derived for time-dependent 
potentials, as expected, the deviations are systematically reduced when we increase level $p$.

\begin{figure}[!b]
\centering
\includegraphics[width=9cm]{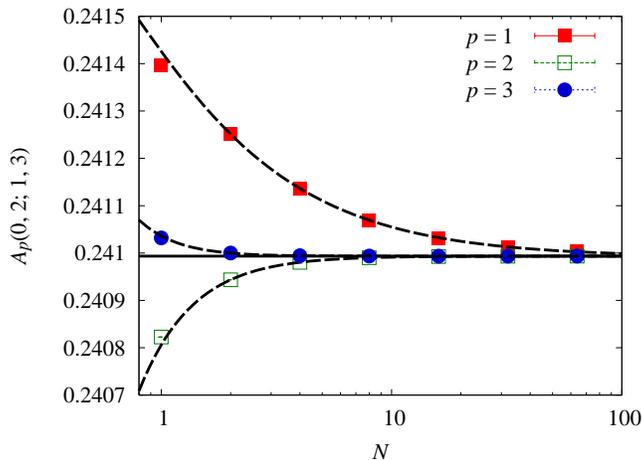}
\caption{Convergence of numerical Monte Carlo  results for the transition amplitude $A_p(0, 2; 1, 3)$ as a function of the number of time steps $N$ 
for the time-dependent harmonic oscillator (\ref{eq:hogrosche}) with  $\omega=1$, calculated with level $p=1, 2, 3$ effective actions. The dashed lines give the fitted functions 
$A_p+B_p/N^p+\ldots$, where the constant term  $A_p$ corresponds to the continuum-theory amplitude $A_p(0, 2; 1, 3)$. The number of MC samples was $N_{\rm MC}=2\cdot 10^{9}$. MC error bars of around $2\times 10^{-8}$ for all values of $p$ are shown in the graph.}
\label{fig:hogrosche-mc}
\end{figure}

To show that the derived short-time effective actions can be used for calculating long-time transition amplitudes using the standard time discretization approach for path integrals,
we display  in 
Fig.~\ref{fig:hogrosche-mc} the convergence of the discretized long-time transition amplitude for the Grosche-rescaled harmonic oscillator as a function of the number of time steps $N$. As 
expected, we have obtained $1/N^p$ behavior, in accordance with our earlier conclusion on the $N$-scaling of errors in this case. The numerical results presented in this graph are obtained using 
the Monte Carlo SPEEDUP \cite{speedup} code, which was modified to include time-dependent effective actions implemented to high orders in the $C$ programming language.

\begin{figure}[!b]
\centering
\includegraphics[width=8.1cm]{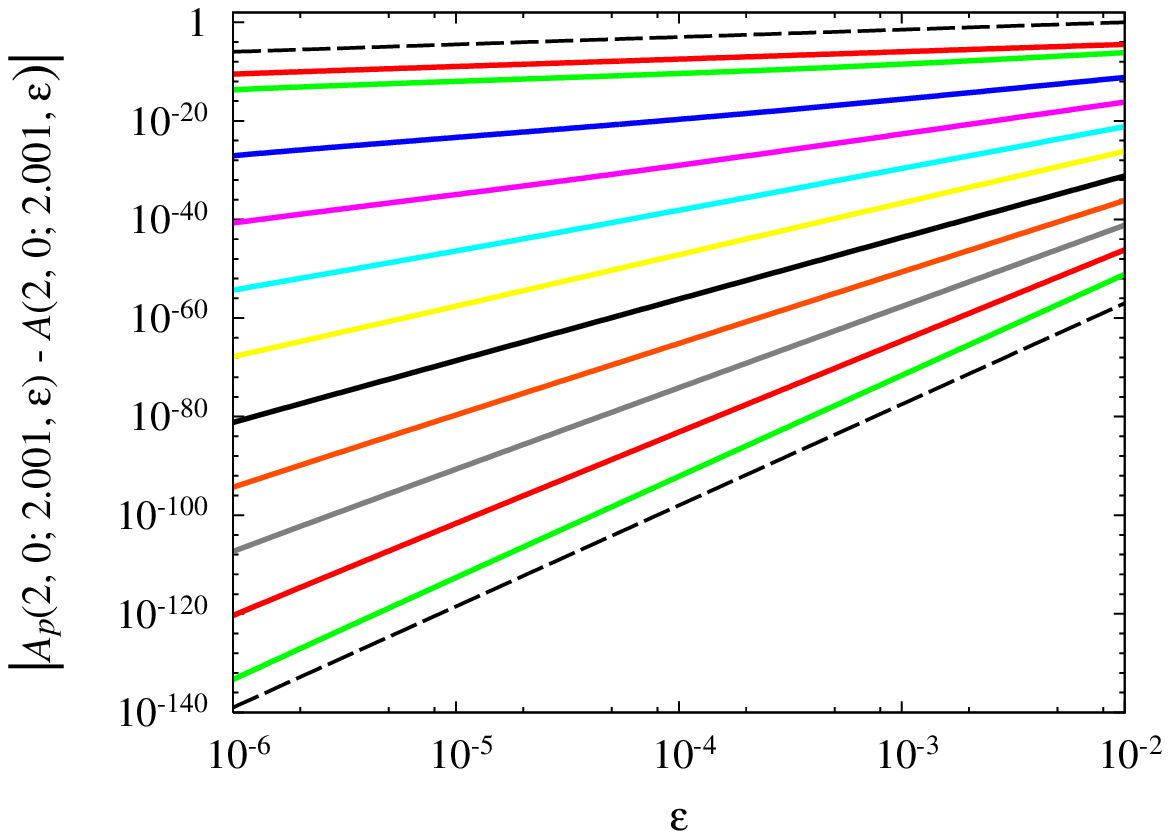}
\includegraphics[width=8.1cm]{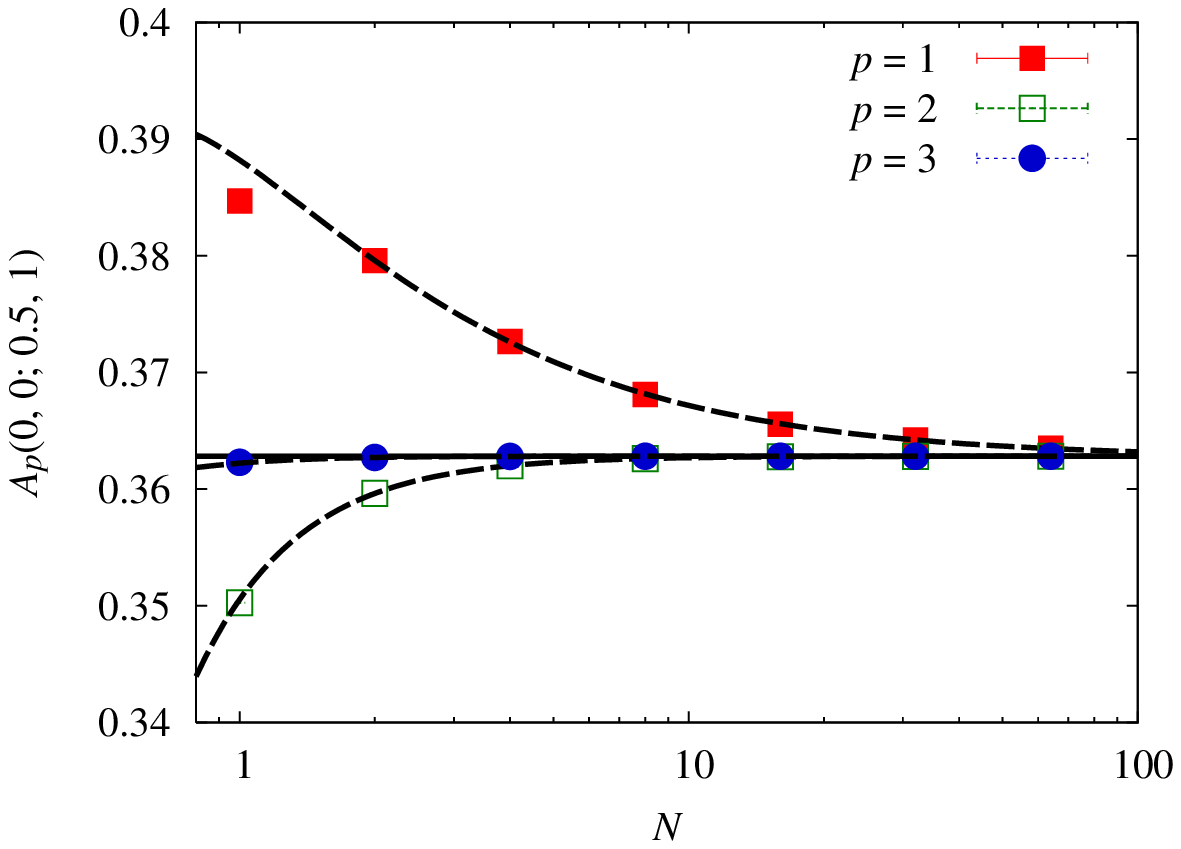}
\caption{(left) Deviations $|A_p(2, 0; 2.001, \varepsilon)-A (2, 0; 2.001, \varepsilon)|$ as a function of propagation time $\varepsilon$ for the forced harmonic 
oscillator (\ref{eq:fho}) with $\omega=\Omega=1$, calculated analytically for $p=1, 2, 4, 6, 8, 10, 12, 14, 16, 18, 20$  from top to bottom. The dashed lines are proportional to $\varepsilon^{1.5}$ and $\varepsilon^{20.5}$, and demonstrate the perfect scaling of the corresponding level $p=1$ and $p=20$ results. (right) Convergence of numerical Monte Carlo  results for 
the transition amplitude $A_p(0, 0; 0.5, 1)$ as a function of the number of time steps $N$ for $p=1, 2, 3$. As before, dashed lines give the fitted functions 
$A_p+B_p/N^p+\ldots$, demonstrating the expected $1/N^p$ scaling of the deviations.  The number of MC samples was $N_{\rm MC}=2\cdot 10^{9}$. MC error bars of around $8\times 10^{-7}$ for all values of $p$ are shown in the graph.}
\label{fig:fho}
\end{figure}

The second exactly solvable model, which we have considered, is the forced harmonic oscillator \cite{feynmanhibbs},
\begin{equation}
\label{eq:fho}
V_{\rm FHO}(x, t)=\frac{1}{2}\omega^2x^2-x\sin\Omega t\, ,
\end{equation}
where $\Omega$ denotes the frequency of the external driving field. Fig.~\ref{fig:fho} presents numerical results for this model for $\omega=\Omega=1$. The top 
plot gives deviations for the case of short-time transition amplitude, calculated analytically using level $p$ effective action. We see again a perfect $\varepsilon$-scaling of deviations. 
The bottom plot shows convergence of a long-time transition amplitude as a function of the number of time steps $N$, illustrating the expected $1/N^p$ behavior.

\begin{figure}[!b]
\centering
\includegraphics[width=8.1cm]{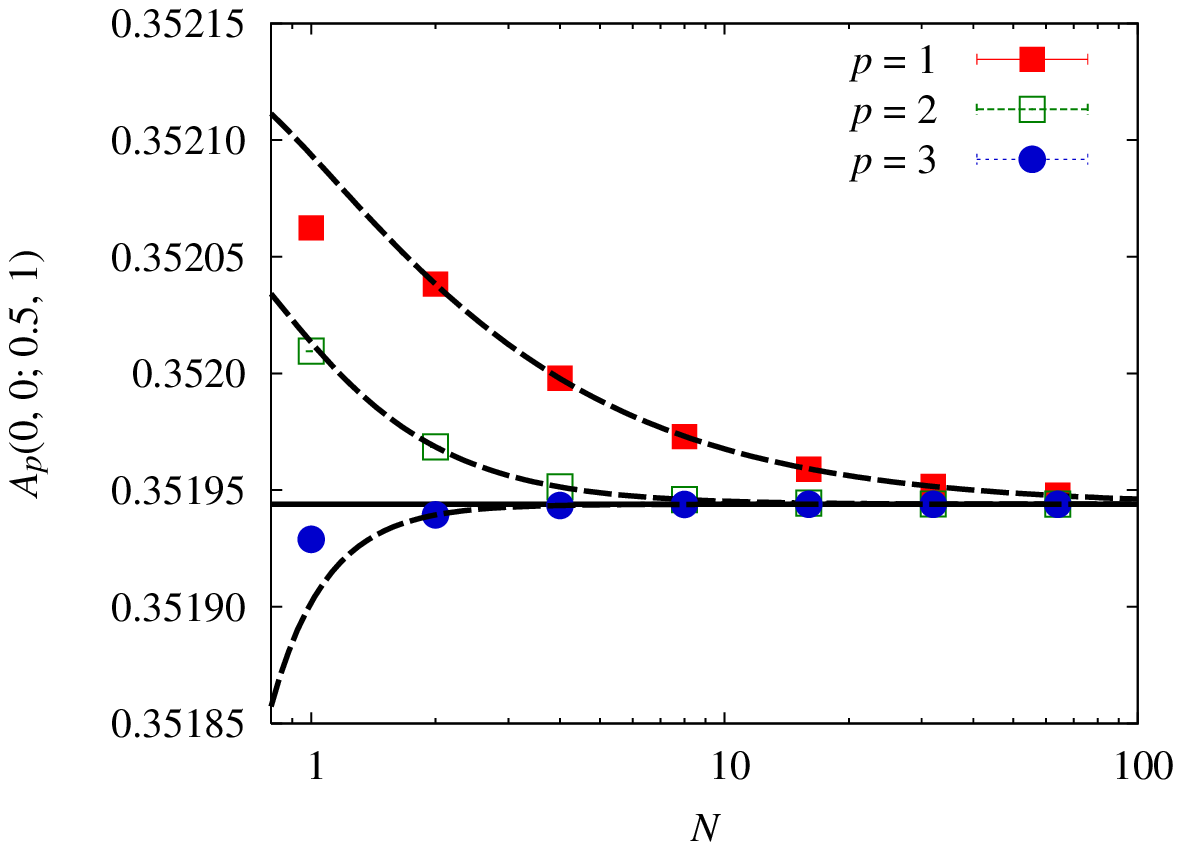}
\includegraphics[width=8.1cm]{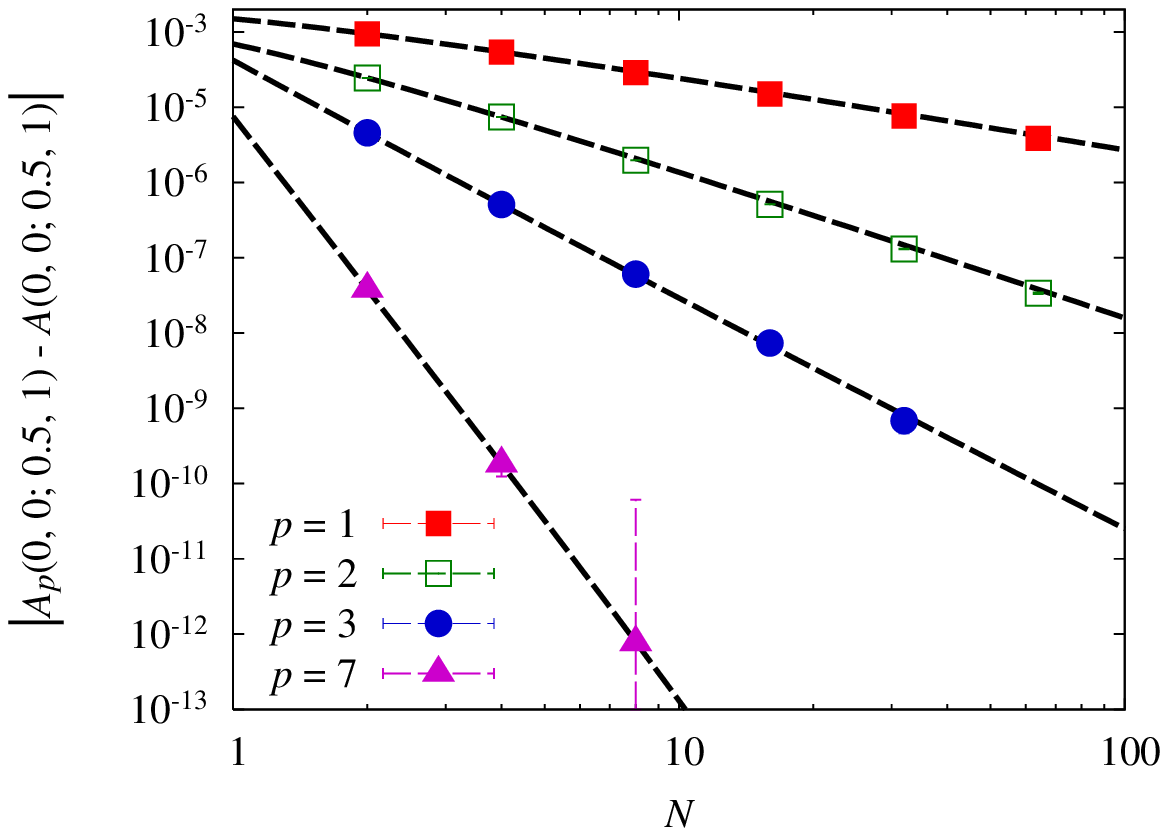}
\caption{(left) Convergence of  numerical  Monte Carlo results for the transition amplitude $A_p(0, 0; 0.5, 1)$ as a function of the number of time steps $N$ for the pure quartic oscillator (\ref{eq:pqgrosche}) with the anharmonicity $g=0.1$, rescaled by the Grosche factor $\zeta (t)=\sqrt{t^2+1}$, 
and calculated with level $p=1, 2, 3$ effective actions. As before, dashed lines give 
the fitted functions $A_p+B_p/N^p+\ldots$. (right) In order to convincingly 
demonstrate the expected dominant $1/N^p$ behavior of deviations, we plot $|A_p(0, 0; 0.5, 1)-A(0, 0; 0.5, 1)|$ 
as a function of the number of time steps $N$ for $p=1,2,3,7$. The dashed lines are fitted polynomials $A_p+B_p/N^p+\ldots$, 
the same as on the top graph. The exact value of the amplitude is obtained 
as the constant term from fitting $p=7$ results. The number of Monte Carlo samples was $N_{\rm MC}=1.6\cdot 10^{13}$ for $p=7$, in order to be able to resolve the exceedingly small deviations from the 
exact value of the amplitude. For $p=1, 2, 3$ we used much smaller values of $N_{\rm MC}$, typically $10^8$ to $10^{10}$. MC error bars of around $2\times 10^{-10}$ for $p=1,2,3$ and  $6\times 10^{-11}$ for $p=7$ are shown in both graphs.}
\label{fig:pqgrosche}
\end{figure}

We have numerically also considered the non-trivial case of a pure quartic oscillator $V_{\rm PQ}(x)=gx^4/24$ rescaled with the same Grosche factor $\zeta(t)=\sqrt{1+t^2}$:
\begin{equation}
\label{eq:pqgrosche}
V_{\rm G, PQ}(x, t)=\frac{gx^4}{24 (1+t^2)^3}\, .
\end{equation}
This model is not exactly solvable, but the continuous transition amplitude
can be determined numerically. To this end one could use the Grosche mapping, while relying on the previous numerical approach \cite{balazpre}, providing 
exact transition amplitudes for the time-independent counterpart of the potential $V_{\rm PQ}$. Another possibility is based on 
numerically obtained results from the modified SPEEDUP code, relying on the fitting of discretized 
amplitudes to the $\varepsilon\to 0$ limit. As we can see from Fig.~\ref{fig:pqgrosche}, the numerical results exhibit a perfect scaling behavior for this non-trivial
model as well. The power-law scaling of 
deviations in the calculated discretized transition amplitudes up to exceedingly small values fully verifies the presented analytic derivation of effective actions.

\begin{figure}[!b]
\centering
\includegraphics[width=9cm]{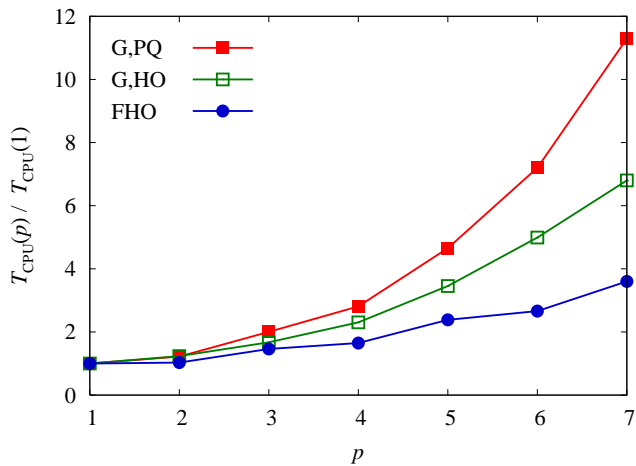}
\caption{Increase in the computational complexity of numerical simulations as a function of level $p$ for the three models considered (Grosche-rescaled harmonic oscillator - G,HO; forced harmonic oscillator - FHO; Grosche-rescaled pure quartic oscillator - G,PQ). The complexity is measured by the relative increase in the CPU time $T_\mathrm{CPU}(p)/T_\mathrm{CPU}(1)$ for calculation of transition amplitudes from Figs.~\ref{fig:hogrosche-mc}, \ref{fig:fho}, and \ref{fig:pqgrosche}, for each model correspondingly. The number of Monte Carlo samples was $N_{\rm MC}=2\cdot 10^{9}$ and number of time steps $N=64$.}
\label{fig:timing}
\end{figure}

As we have demonstrated during our analysis of several examples, the main advantage of the effective action approach is the calculation of transition amplitudes with high precision, which can be improved by using higher levels $p$. In practical applications, this increase in the precision is counterweighted by having to evaluate increasingly complex expression for the effective action. Fig.~\ref{fig:timing} gives the comparison of the increase in the computational complexity of numerical codes for different levels $p$ for the three models considered. As we can see, depending on the model, CPU time increases 5 to 10 times for $p=7$, and clearly the benefit of 7 orders of magnitude increase in the precision of calculations far outweighs the CPU time increase. However, as $p$ increases, the computational complexity will become too high. Since this strongly depends on the model, one has to perform a small-scale complexity study, similar to the one presented in Fig.~\ref{fig:timing},  in order to find optimal value of level $p$ to be used in numerical simulations.

While the presented approach is excellently suited for applications where high-precision transition
amplitudes are necessary, e.g.~the calculation of partition functions, other 
higher-order schemes may be preferred for solving different types of problems. For example, the split-operator method \cite{chinchen, omelyanpre, omelyan, bayepre, chinkrotscheck, k1} is ideally suited 
for studying both the real- and the imaginary-time 
evolution of various quantum systems. Furthermore, it has the advantage that it can be implemented in any chosen representation, which is appropriate for the quantum system, while the effective 
action approach relies on using the coordinate representation.

\section{Conclusions}
\label{sec:conclusion}
We have presented an analytic procedure for deriving the short-time expansion of the propagator for a general one-particle non-relativistic quantum system in a time-dependent potential up to previously inaccessible high orders. The procedure is based on recursively solving both the forward and backward Schr\" odinger equation for the transition amplitude. Following an earlier approach for time-independent potentials \cite{balazpre}, we have derived  recursion relations which allow an efficient analytic calculation of the effective potential to arbitrarily high orders in the propagation time $\varepsilon$ in the imaginary time. The analytically derived results are numerically verified by studying several simple models.

The presented approach will be further expanded and generalized to many-body and multi-component systems in the forthcoming publication \cite{fcpitdp2}. The next paper will also include generalization of the present imaginary-time approach to the real-time formalism, and its extensive numerical verification.

\section*{Acknowledgements}
We thank Hagen Kleinert for several useful suggestions.
This work was supported in part by the Ministry of Science and Technological Development of the Republic of Serbia, under project No. ON171017 and bilateral projects PI-BEC and NAD-BEC, funded jointly with the German Academic Exchange Service (DAAD). Numerical simulations were run on the AEGIS e-Infrastructure, supported in part by FP7 project EGI-InSPIRE, HP-SEE and PRACE-1IP.

%------------------------------------------------------------------------------
\end{document}